\documentclass[twoside,a4paper,11pt]{proceedings}
\usepackage{graphicx}
\usepackage{hyperref}
\usepackage{movie15}
\usepackage{natbib}
\begin{document}
\pagenumbering{arabic}
\pagestyle{myheadings}
\thispagestyle{empty}
\vspace*{-1cm}
{\flushleft\includegraphics[width=8cm,viewport=0 -30 200 -20]{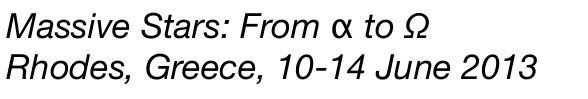}}
\vspace*{0.2cm}
\begin{flushleft}
{\bf {\LARGE
Rotational properties of the O-type star population in the Tarantula region
}\\
\vspace*{1cm}
O.H. Ram\'irez-Agudelo$^1$,
S. Sim\'on-D\'{i}az $^{2,3}$,
H. Sana$^{1,4}$,
A. de Koter$^{1,5}$, \\
C. Sab\'{i}n-Sanjul\'{i}an$^{2,3}$,
S.E. de Mink$^{6,7}$, 
P. L. Dufton$^{8}$,
G. Gr\"afener$^{9}$, \\
C.J. Evans$^{10}$,
A. Herrero$^{2,3}$,
N. Langer$^{11}$,
D.J. Lennon$^{12}$,
J. Ma\'{i}z Apell\'aniz$^{13}$, \\
N. Markova$^{14}$,
F. Najarro$^{15}$,
J. Puls$^{16}$,
W.D. Taylor$^{10}$,
and
J.S. Vink$^{9}$ 
%
}\\
\center
\textit{(Affiliations can be found after the references)}

\vspace*{0.5cm}
%

%
\end{flushleft}
\markboth{
Rotational properties of the O-type star population in the Tarantula region
}{
Ram\'irez-Agudelo et al.
}
\thispagestyle{empty}
\vspace*{0.4cm}
\begin{minipage}[l]{0.09\textwidth}
\ 
\end{minipage}
\begin{minipage}[r]{0.9\textwidth}
\vspace{1cm}
\section*{Abstract}{\small
The 30 Doradus (30\,Dor) region in the Large Magellanic Cloud (also known as the Tarantula Nebula) is the nearest massive starburst region, containing the richest sample of 
massive stars in the Local Group. It is the best possible laboratory to investigate aspects of the formation and evolution of massive stars. Here, we focus on rotation which is a 
key parameter in the evolution of these objects. We establish the projected rotational velocity, $v_{e}\sin i$, distribution of an unprecedented sample of 216 radial 
velocity constant ($\rm{\Delta RV\, \leq\, 20 \,km s^{-1}}$) O-type stars in 30\,Dor observed in the framework of the VLT-FLAMES Tarantula Survey (VFTS). The distribution 
of $v_{e}\sin i$ shows a two-component structure: a peak around 80 $\rm{km s^{-1}}$ and a high-velocity tail extending up to $\sim$600 $\rm{km s^{-1}}$. Around 75\% 
of the sample  has 0 $\leq\, v_{e}\sin i \leq$ 200 $\rm{km s^{-1}}$ with the other 25\% distributed in the high-velocity tail. 
The presence of the low-velocity peak is consistent with that found in other studies of late-O and early-B stars.
The high-velocity tail is compatible with expectations from binary interaction synthesis models and may be predominantly populated by post-binary interaction, spun-up, objects and mergers. This may have important implications for the nature of progenitors of long-duration gamma ray bursts.
\vspace{10mm}
\normalsize}
\end{minipage}

\section{Introduction}


$\,\!$\indent Rotation is a key parameter in the evolution of massive stars affecting their evolution, chemical yields, budget of ionizing photons
and their final fate as supernovae and long gamma-ray bursts. The 30\,Dor starburst region in the Large Magellanic Cloud
contains the richest sample of massive stars in the Local 
Group and is the best possible laboratory to investigate aspects of the formation and evolution of massive stars, and to establish statistically meaningful
distributions of their physical properties. In this paper, we report on the measured projected rotational velocity, $v_{e}\sin i$, for more than
200 O-type stars observed as part of the VLT-FLAMES Tarantula Survey \citep[VFTS,][]{evans}.

\section{Sample and Method}
$\,\!$\indent VFTS is a multi-epoch intermediate-resolution spectroscopic campaign 
targeting over 350 O and over 400 early B-type stars across the 30\,Dor region.
The sample used in this paper is composed of 172 O-type stars with no significant radial velocity variations
and 44 which display only small shifts  \citep[$\rm{\Delta RV\, \leq\, 20 \,km s^{-1}}$;][]{sana}.
We use Fourier transform \citep{gray,simon} and line profile fitting \citep{simon1} methods to measure projected rotational velocities.
Discussion of the methods and achieved precision of the full list of measurements can be found in \citet{ramirezagudelo1}

\section{Results and discussion}

$\,\!$\indent The distribution of projected rotational velocities of our sample shows a two-component structure: a low-velocity peak at around 80 $\rm{km s^{-1}}$ and a high velocity-tail starting at about 200 kms$^{-1}$ and extending up to $\sim$600 $\rm{km s^{-1}}$ (see Fig.~\ref{fig:Fig1}). The main conclusions regarding these two components are:

\begin{figure}
\center
\includegraphics[scale=0.45]{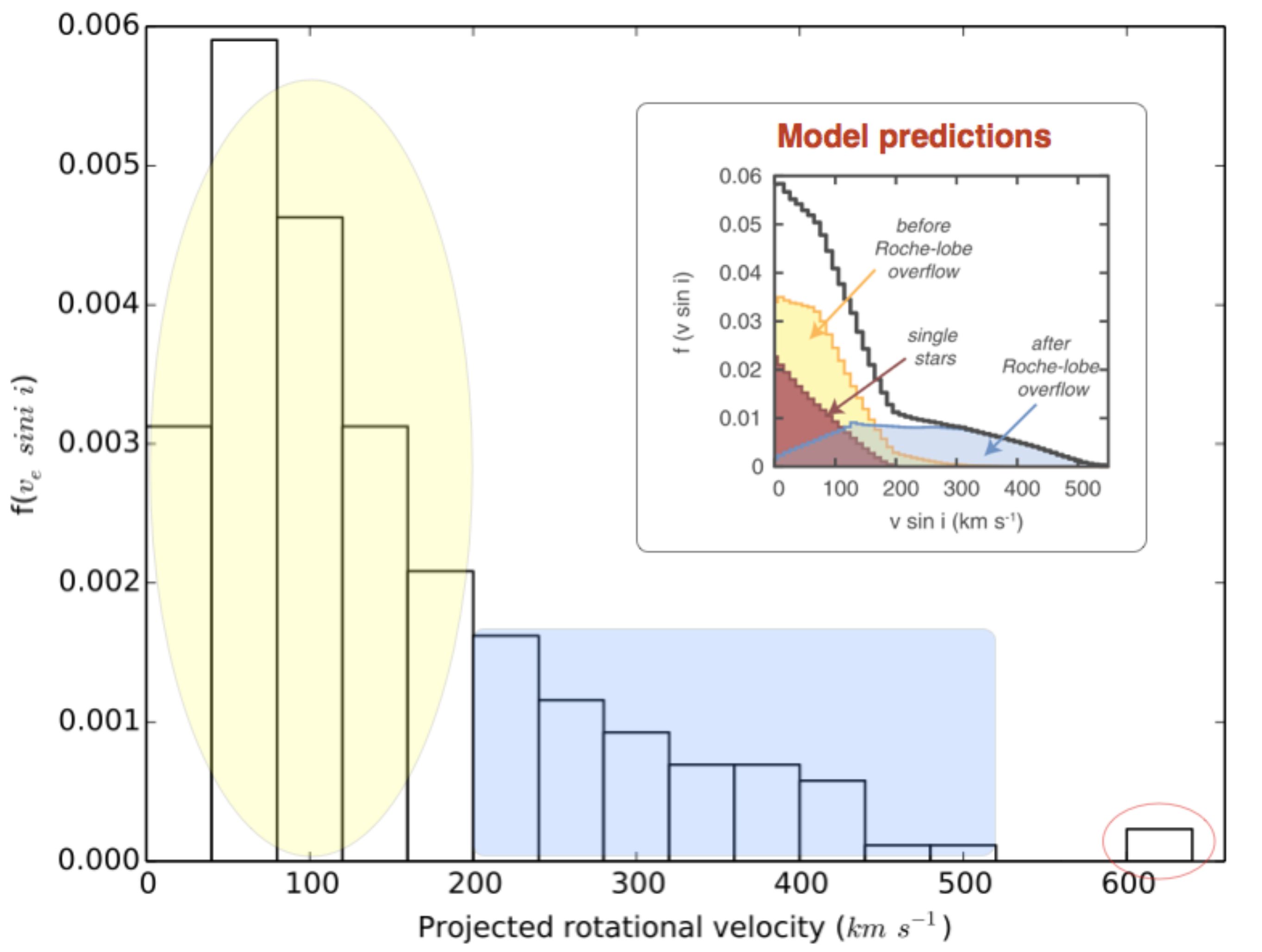} 
\caption{Distribution of the projected rotational velocities of the 216 O-type stars in our 30\,Dor sample. The insert panel shows the
effects of binary interaction on the $\rm{v_{e}\sin i}$ distribution \citep{selma}.}
\label{fig:Fig1}
\end{figure}

\textbf{Slow rotators:} 75\% of our sample stars have $v_{e}\sin i$ $\rm{<}$ 200 $\rm{km s^{-1}}$. These stars thus rotate at less than 20\% of their break-up velocity. For the bulk of the sample, mass loss in a stellar wind and/or envelope expansion is not efficient enough to significantly spin down these stars. If massive-star formation results in fast rotating stars at birth, as one may think from angular momentum conservation considerations, 
most massive stars have to spin down quickly by some other mechanism to reproduce the observed distribution.

\textbf{Fast rotators:} The presence of a well populated high-velocity tail (25\% with $v_{e}\sin i$ $>$ 200 $\rm{km s^{-1}}$) is compatible with predictions from binary evolution. \citet[][see the insert panel in Fig.~\ref{fig:Fig1}]{selma} show that such a tail in the $v_{e}\sin i$ distribution arises naturally from mass transfer and mergers.  A sizable fraction of the post-interaction systems may have only small radial velocity shifts. Combined with the merged systems, the bulk of this population may thus fulfill our selection criterium $\Delta {\rm RV} \leq 20$\,kms$^{-1}$.  That rapid rotators result from spin-up through mass transfer and mergers has important implication for the evolutionary origin of the progenitors of long gamma-ray bursts.  These progenitor systems may be dominated by, or be exclusively due to, post-interacting binaries or mergers.


\section{Conclusion}

We have estimated projected rotational velocities for the presumably single O-type stars in the VFTS sample (216 stars). 
The most distinctive feature of this distribution is a two-component structure with a peak at low velocities and a pronounced tail
at high velocities with over 50 stars rotating faster than 200 $\rm{km s^{-1}}$.  A further discussion of our results, including an analysis of the intrinsic distribution corrected for projection effects and the $v_{e} \sin i$ distribution as a function of spatial distribution, luminosity class and spectral type, is presented in \citet{ramirezagudelo1}.



\small  
%
%

\bibliographystyle{aj}
\small

\newpage
\section*{Affiliations:}
$^{1}$
Astronomical Institute Anton Pannekoek, University of Amsterdam, The Netherlands \\
$^{2}$
Instituto de Astrof\'{i}sica de Canarias, C/ V\'{i}a L\'{a}ctea s/n, E-38200 La Laguna, Tenerife, Spain \\
$^{3}$
Departamento de Astrof\'{i}sica,  Universidad de La Laguna,  Avda. Astrof\'{i}sico Francisco S\'{a}nchez s/n, E-38071 La Laguna, Tenerife, Spain\\
$^{4}$
           Space Telescope Science Institute,
           3700 San Martin Drive,
           Baltimore,
           MD 21218,
           USA\\            
$^{5}$
           Instituut voor Sterrenkunde, 
           Universiteit Leuven, 
           Celestijnenlaan 200 D, 
           3001, Leuven, Belgium \\
$^{6}$
			Observatories of the Carnegie Institution for Science, 
			813 Santa Barbara St, 
			Pasadena, 
			CA 91101, 
			USA\\
$^{7}$
	Cahill Center for Astrophysics, 
	California Institute of Technology, 
	Pasadena, 
	CA 91125, 
	USA\\
$^{8}$
	        Astrophysics Research Centre, 
	        School of Mathematics and Physics, 
	        Queen's University of Belfast, 
	        Belfast BT7 1NN, 
	        UK \\
$^{9}$
           Armagh Observatory,
           College Hill,
           Armagh, BT61 9DG,
           Northern Ireland,
           UK \\
$^{10}$
           UK Astronomy Technology Centre,
           Royal Observatory Edinburgh,
           Blackford Hill, Edinburgh, EH9 3HJ, UK \\
$^{11}$
          Argelander-Institut f\"ur Astronomie, 
           Universit\"at Bonn, 
           Auf dem H\"ugel 71, 
           53121 Bonn, Germany \\
$^{12}$
           European Space Astronomy Centre (ESAC),
           Camino bajo del Castillo, s/n
           Urbanizacion Villafranca del Castillo,
           Villanueva de la Ca\~nada,
           E-28692 Madrid, Spain \\
$^{13}$
           Instituto de Astrof\'{i}sica de Andaluc\'{i}a-CSIC,
           Glorieta de la Astronom\'ia s/n,
           E-18008 Granada, Spain \\
$^{14}$
			Institute of Astronomy with NAO,
			Bulgarian Academy of Science,
			PO Box 136,
			4700 Smoljan,
			Bulgaria \\
$^{15}$
			Centro de Astrobiolog\'{i}a (CSIC-INTA),
			Ctra. de Torrej\'on a Ajalvir km-4,
			E-28850 Torrej\'on de Ardoz,
			Madrid,
			Spain \\
$^{16}$
			Universit\"atssternwarte
			Scheinerstrasse 1,
			81679 M\"unchen, 
			Germany
\end{document}